\documentclass[prl, twocolumn, amsmath, amssymb, aps, floatfix]{revtex4-2}

\usepackage{graphicx}
\usepackage{dcolumn}
\usepackage{bm}
\usepackage[bookmarks=false]{hyperref}
\usepackage{nicefrac}
\hypersetup{
	colorlinks   = true, 
	urlcolor     = blue, 
	linkcolor    = blue, 
	citecolor   = blue 
}
\usepackage{epstopdf}
\usepackage{braket}
\usepackage{xcolor}
\usepackage[normalem]{ulem}

\begin{document}

\title{Excited-State Phase Diagram of a Ferromagnetic Quantum Gas}

\author{B. Meyer-Hoppe$^1$}
\email[Author to whom correspondence should be addressed: ]{meyer-hoppe@iqo.uni-hannover.de}
\author{F. Anders$^1$} 
\author{P. Feldmann$^{2,3,4}$}
\author{L. Santos$^2$}
\author{C. Klempt$^{1}$}

\affiliation{$^1$Leibniz Universit\"at Hannover, Institut für Quantenoptik, Welfengarten 1, D-30167 Hannover, Germany  \\
$^2$Leibniz Universit\"at Hannover, Institut f\"ur Theoretische Physik, Appelstra\ss{}e~2, D-30167~Hannover, Germany\\ 
$^3$Stewart Blusson Quantum Matter Institute, The University of British Columbia, 2355 East Mall, Vancouver, British Columbia, V6T 1Z4, Canada\\
$^4$Department of Physics and Astronomy, The University of British Columbia, 6224 Agricultural Road, Vancouver, British Columbia, V6T 1Z1, Canada}

\begin{abstract}

The ground-state phases of a quantum many-body system are characterized by an order parameter, which changes abruptly at quantum phase transitions when an external control parameter is varied. Interestingly, these concepts may be extended to excited states, for which it is possible to define equivalent excited-state quantum phase transitions. However, the experimental mapping of a phase diagram of excited quantum states has not yet been realized. Here we present the experimental determination of the excited-state phase diagram of an atomic ferromagnetic quantum gas, where, crucially, the excitation energy is one of the control parameters. The obtained phase diagram exemplifies how the extensive Hilbert state of quantum many-body systems can be structured by the measurement of well-defined order parameters.

\end{abstract}

\maketitle
Last century's experimental advancement to cool quantum systems close to absolute zero temperature, where thermal fluctuations are frozen out, has led to a revolution in the understanding of quantum phases and phase transitions~\cite{Sachdev2011}.
Nowadays, quantum systems may be shielded from the surrounding environment to prevent thermalization.
As a result, it is possible to investigate the properties and dynamics of quantum systems in states different than the ground state.
Major theoretical efforts, followed by experimental demonstrations, have focused on extending the concept of quantum phases and phase transitions to the realm of excited states.
This includes dynamical systems, where phase transitions can be associated with a sudden change of the long-term average of observables~\cite{Heyl2013,Jurcevic2017,Zunkovic2018,Smale2019,Chu2020,Muniz2020}.
The dynamics can also result in a sudden change of oscillatory behavior, termed time crystal~\cite{Zhang2017,Choi2017}, and can show universal features~\cite{Ritsch2013,Prufer2018,Erne2018}. 
Furthermore, in open quantum systems, the interplay between driving and dissipation may lead to dissipative phase transitions, characterized by a sudden change of the steady-state properties~\cite{Kessler2012}.

Interestingly, isolated quantum systems may present so-called excited-state quantum phase transitions~({ESQPTs}).
At these transitions, occurring at a particular excitation energy, there is a qualitative change in the nature of the excited states, which hence form well differentiated excited-state quantum phases~\cite{Cejnar2021}.
The presence of ESQPTs, with their characteristic divergence of the density of states, has been experimentally revealed in microwave Dirac billiards~\cite{Dietz2013} and molecular spectra~\cite{Winnewisser2005,Zobov2005,Larese2013, Khalouf-Rivera2021}. 
However, despite some theoretical proposals that have identified possible order parameters in some scenarios~\cite{Brandes2013,Puebla2013,Puebla2014,Feldmann2021,Cabedo2021}, the experimental mapping of an excited-state phase diagram by measuring appropriate order parameters remains an open challenge in any physical platform.

\begin{figure}[ht]
\centering
  \includegraphics[width=\columnwidth]{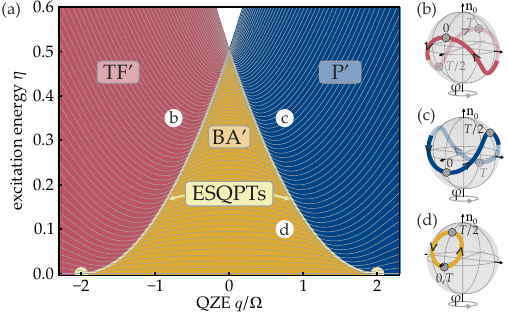}
    \caption{
  \label{fig:abstract}
  Excited-state quantum phase diagram.
  (a)~Three excited-state quantum phases TF$^{\prime}$ (twin-Fock-like; red, left), BA$^{\prime}$ (broken-axisymmetric-like; orange, center), and P$^{\prime}$ (polar-like; blue, right) appear for two control parameters, the excitation energy per particle $\eta$ and the quadratic Zeeman energy $q$.
  Gray lines represent the eigenenergies of Eq.~\eqref{eq:HamiltonianQ} calculated for $N=70\,000$ atoms, where every 500th eigenvalue is plotted.
  The excited-state quantum phase transitions (ESQPTs) are indicated by a light yellow line and the ground-state quantum phase transitions are highlighted by light yellow dots.
  (b)-(d) Bloch-sphere trajectories for control parameters indicated in (a).
  }
\end{figure}
\begin{figure}[ht]
\centering
  \includegraphics[width=\columnwidth]{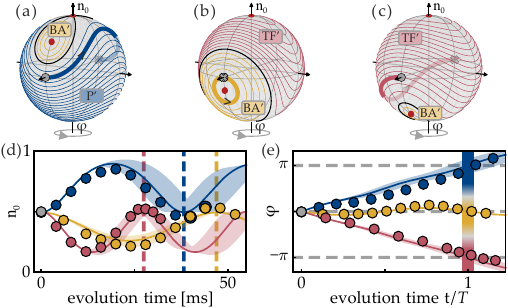}
    \caption{
  \label{fig:stepwise}
  Interferometric phase as an order parameter.
  (a)-(c)~Depending on the QZE $q/\Omega=\{1.25, -0.5, -1.5\}$, the quantum states follow trajectories with different topologies that can be associated with the P$^{\prime}$ (blue), BA$^{\prime}$ (orange), and TF$^{\prime}$ (red) quantum phases with a separatrix in between (black lines).
  (d) Measurement of the relative population $n_0$ (circles) as a function of evolution time corresponding to the highlighted trajectories in (a)-(c).
  The data are recorded by an iterative measurement with reinitialization at each data point ($4-6$~ms), as described in the Supplemental Material~\cite{supp}.
  Statistical error bars are smaller than the symbols.
  Solid lines and shading represent the theoretical prediction with a systematical uncertainty in $q/\Omega$.
  The dashed lines mark a full population oscillation period $T$.
  (e) A $\pi/2$ rotation on the Bloch sphere by an RF coupling pulse allows for a measurement of the phase, which represents an order parameter for ESQPTs (color bar).
  }
\end{figure}

Coherent spin dynamics has been studied in different ultracold-gases scenarios, including experiments on external~\cite{Albiez2005} and internal~\cite{Zibold2010} Josephson oscillations, where distinct dynamical regimes were observed depending on the state initialization. 
Coherent spin dynamics has been also experimentally studied in an atomic spinor Bose-Einstein condensate (sBEC)~\cite{Chang2005}, where also different dynamical regimes are expected as a function of the initial conditions~\cite{Zhang2005, Kawaguchi2012}. 
Interestingly, these different dynamical regimes could be linked with the idea of excited-state phases~\cite{Feldmann2021}. 
However, as for other physical systems, also for ultracold gases the experimental mapping of the corresponding excited-state phase diagram has not been achieved so far.

In this Letter, we experimentally map out the excited-state phase diagram of a ferromagnetic $F=1$ sBEC. 
In this system, ground-state phase transitions may be driven by the variation of the quadratic Zeeman energy $q$, which acts as the control parameter~\cite{Kawaguchi2012, Stamper-Kurn2013, Hoang2016,Zou2018}.
Recent experiments, performed with a sodium sBEC~\cite{Tian2020}, have revealed a modification of the spin dynamics closely associated with an ESQPT, but were limited to changing $q$ of the highest-energy level.
In contrast to ground-state (or highest-state) transitions, general ESQPTs can be crossed not only by varying $q$, but also as a function of the excitation energy, which serves as a second control parameter. 
This is achieved in our experiments by creating coherent-state spin superpositions with an energy that can be carefully adjusted by both the population of the spin states and their relative phase. 
We characterize each point of the diagram by means of an interferometric order parameter inspired by Ref.~\cite{Feldmann2021}, but robust with respect to magnetic field noise.
With the help of this order parameter, we obtain the complete excited-state phase diagram as a function of the two control parameters, clearly identifying the three distinct excited-state quantum phases.

We initially prepare an sBEC of $7\times10^4$ rubidium atoms in the hyperfine state $|F,m \rangle = |1,0\rangle$ in a crossed-beam optical dipole trap.
The spin dynamics, characterized by the creation and annihilation of pairs of atoms in $|1,\pm 1 \rangle$, is modeled by the Hamiltonian in single-mode approximation~\cite{Kawaguchi2012}
\begin{align}
    \label{eq:HamiltonianQ}
    \hat H = ~&q\left(\hat N_{+1}+\hat N_{-1}\right) \notag
    - \frac{\Omega}{N} \left[ \left(\hat N_0 -\frac{1}{2} \right)\left( \hat N_{+1}+\hat N_{-1} \right) \right.\\
    & \left. +
    \hat a_1^\dagger \hat a_{-1}^\dagger \hat a_0 \hat a_0+\hat a_0^{\dagger}\hat a_0^{\dagger}\hat a_1\hat a_{-1}
    \right]\!,
\end{align}
where  $\hat a_m^\dagger$ and $\hat a_m$ are the bosonic creation and annihilation operators for state $m$, and $\hat N_m\equiv \hat a^\dagger_m \hat a_m$ with $\sum_m \hat N_m=N$.
The interaction strength $\Omega=h \times 13.9\,$Hz is experimentally determined and depends on the atom number, the spatial wave function and the atomic properties.
Note that we assume the magnetization-free subspace, $\langle N_{+1}-N_{-1} \rangle = 0$, which eliminates the influence of the linear Zeeman effect.
The quadratic Zeeman energy (QZE) $q$ is initially positive, but can be adjusted to positive and negative values by a microwave dressing field~\cite{Anders2021}.

The Hamiltonian~\eqref{eq:HamiltonianQ} features three ground-state phases~\cite{Kawaguchi2012,Luo2017} depending on the QZE: (i) the twin-Fock~(TF) phase for $q/\Omega<-2$, (ii) the polar~(P) phase for $q/\Omega>2$, and (iii) the intermediate broken-axisymmetry~(BA) phase for $|q/\Omega|<2$.
Figure~\ref{fig:abstract}(a) shows the energy of the ground state and a series of exemplary excited states, as obtained by exact diagonalization from Eq.~\eqref{eq:HamiltonianQ}~\cite{Feldmann2021}.
The vanishing gap between the ground and first excited state at $q/\Omega=\pm 2$ marks the ground-state phase transitions.

A vanishing gap between adjacent energy eigenstates persists also at increasing excitation energy per particle $\eta=\langle \hat H\rangle/\left(\Omega N\right) - \eta_0 $ with the ground state energy $\eta_0$, but shifts towards smaller $|q/\Omega|$.
This diverging density of states marks the ESQPTs, which separate the excited-state spectrum into three qualitatively different excited-state phases.
In analogy with the ground-state phase labels, we denote these phases as twin-Fock-like~(TF$^{\prime}$), broken-axisymmetric-like~(BA$^{\prime}$), and polar-like~(P$^{\prime}$).
Note that contrary to the ground-state transitions, or the equivalent transition in the most energetic excited state~\cite{Tian2020}, the ESQPTs can be crossed not only by quenching $q$ [Figs.~\ref{fig:abstract}(b)-(c)], but, crucially, also by a controlled change of the excitation energy $\eta$ at a fixed $q$ value [Figs.~\ref{fig:abstract}(c)-(d)].

The main purpose of this work is to experimentally determine the three excited-state phases at an arbitrary excitation energy.
This requires, in addition to the introduction of an appropriate robust order parameter discussed below, the capability of preparing states with a controllable and well-defined nonzero energy.
The sBEC is initialized in the state $|1,0 \rangle$ at the desired QZE $q$ by a sudden quench of the microwave dressing field.
Subsequently, the energy is set by a variable population transfer and phase adjustment which generate a coherent spin state.
Resonant radio-frequency (RF) radiation couples the level $|1,0 \rangle$ to the the symmetric superposition $|g \rangle \equiv \frac{1}{\sqrt{2}} (|1,1 \rangle+|1,-1 \rangle)$ with a Rabi frequency $\Omega_R$.
The RF pulse duration $\tau$ adjusts a relative population $n_0(\tau)=\textrm{cos}^2(\frac{1}{2}\Omega_R \tau)$ in level $|1,0 \rangle$.
The resulting state can be visualized on the generalized Bloch sphere for the two levels $|1,0 \rangle$ and $\ket{g}$.
The antisymmetric superposition $|h \rangle \equiv \frac{1}{\sqrt{2}} (|1,1 \rangle-|1,-1 \rangle)$ is not populated and therefore remains negligible under the action of Eq.~\eqref{eq:HamiltonianQ}.
The relative phase difference $\varphi$ between $|g \rangle$ and $|1,0 \rangle$, i.e., the azimuthal angle of the Bloch sphere, can be adjusted from its reference value $\varphi=-\pi/2$ directly after the RF pulse to any chosen value by an off-resonant microwave pulse addressing the $|1,0 \rangle \leftrightarrow |2,0 \rangle$ transition. The created state is not a single, stationary eigenstate, but a superposition of excited eigenstates within a narrow energy window, which allows sampling the phase diagram with high resolution.

After preparation, spin-changing collisions result in a time evolution according to Eq.~\eqref{eq:HamiltonianQ}, as visualized on the Bloch spheres in Figs.~\ref{fig:abstract}(b)-(d).
The different excited-state quantum phases are characterized by trajectories of differing topology.
Besides the fixed points, the dynamics always leads to an oscillation of the population with variable amplitude, while the phase evolution is either bounded (BA$^{\prime}$ phase) or stretches over a full $2\pi$ rotation, either in positive (P$^{\prime}$) or negative (TF$^{\prime}$) direction.
If the initial state is prepared at $n_0=0.5$ and $\varphi=0$, the state evolves according to the trajectories in Figs.~\ref{fig:stepwise}(a)-(c).
The highlighted association with the excited-state quantum phase can be determined from a measurement of the azimuthal phase after a full oscillation cycle of the population, which is $\pi$, 0, and $-\pi$ for the P$^{\prime}$, BA$^{\prime}$, and TF$^{\prime}$ phases~\cite{Feldmann2021}.

\begin{figure}[t!]
\centering
  \includegraphics[width=\columnwidth]{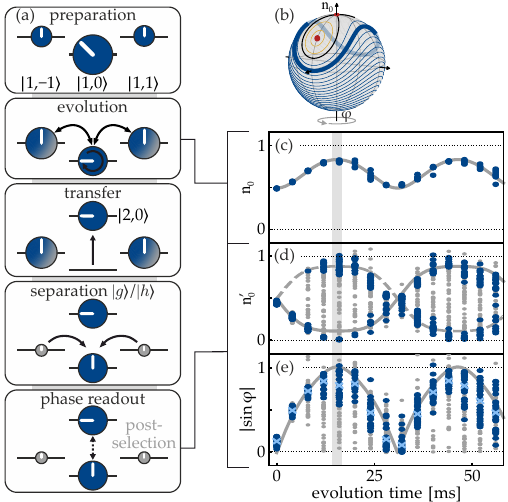}
    \caption{
  \label{fig:sequence}
  Measurement of an improved interferometric order parameter.
  (a) Illustration of a sequence to measure the phase of the atoms in $|g \rangle$ with respect to $|1,0 \rangle$, while using the atoms in $|h \rangle$ for postselection to reduce magnetic-field sensitivity.
  (b) An examined trajectory in the P$^{\prime}$ phase.
  (c)-(e)~Measurement of the relative population $n_0$ (projection onto vertical axis in (b)), the relative population for phase readout $n'_0$ (horizontal axis), and the order parameter $|\sin\varphi|$.
  Measurement results excluded by the postselection are marked in gray.
  Light blue crosses indicate mean values of the remaining measurement data (dark blue) for $|\sin\varphi|$.
  Solid lines are obtained from the $N\rightarrow\infty$ limit of Eq.~\eqref{eq:HamiltonianQ}, dashed lines include an additional phase of $\pi$ on $|g \rangle$, the shaded area indicates half a population oscillation $T/2$.
  }
\end{figure}

\begin{figure*}[t]
\centering
  \includegraphics[width=\textwidth]{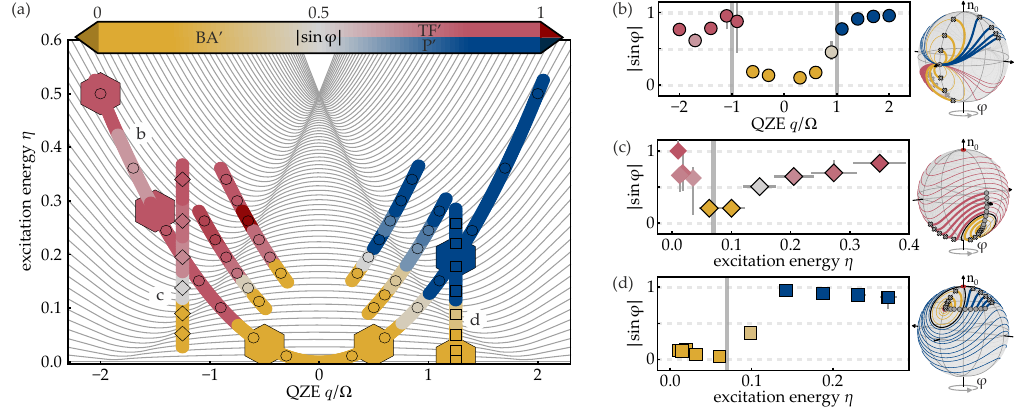}
    \caption{
  \label{fig:orderparameterdiagram}
  Measurement of the order parameter as a function of various control parameters.
  (a) For the description of the phase diagram, see Fig.~\ref{fig:abstract}.
  The recorded order parameter $|\sin\varphi|$ (color scale) is recorded along the colored lines, enabling a clear determination of the quantum phases.
  Small circles correspond to a variation of the QZE $q$, and small diamonds and squares result from a variation of the excitation energy $\eta$, either by adjusting the population or the phase, respectively.
  The large hexagons represent an iterative measurement of an equivalent phase-dependent order parameter (see Fig.~\ref{fig:stepwise}).
  (b) A variation of the QZE $q$ leads to qualitatively different Bloch-sphere trajectories.
  The order parameter (circles) after $T/2$ distinguishes between the quantum phases TF$^{\prime}$, BA$^{\prime}$, and P$^{\prime}$.
  (c) Equivalent measurement for a variation of the relative population $n_0$.
  The gray circles on the Bloch sphere indicate different starting points for the trajectories.
  (d) The order parameter as a function of the initial phase $\varphi$ is evaluated at $T/4$ (see text). 
  The Bloch sphere is displayed from behind.
  In (b)-(d), vertical error bars indicate the sensitivity to the postselection.
  Horizontal error bars are obtained from measured fluctuations of $n_0$, $\varphi$, and $q/\Omega$.
  The expected phase transitions are depicted by gray vertical lines.
  All colors reflect the order parameter, as employed for (a).
  }
\end{figure*}

A direct measurement of this phase by an RF pulse is hindered by the detrimental effect of magnetic-field fluctuations~\cite{Cabedo2021}, in our case {$\Delta B = 47\,\mu{}$G} from shot to shot, which couple the symmetric level $|g \rangle$ with the antisymmetric level $|h \rangle$ and lead to a dephasing after $5-10$~ms [Fig.~\ref{fig:sequence}(d)].
We thus employ a protocol that is largely insensitive to magnetic field fluctuations [see Fig.~\ref{fig:sequence}(a) and \cite{supp}].
The phase measurement protocol starts after the evolution along the trajectory by transferring the atoms in level $|1,0 \rangle$ to $|2,0 \rangle$.
A subsequent radio-frequency $\pi$ pulse transfers all atoms from level $|g \rangle$ to $|1,0 \rangle$, while leaving all $|h \rangle$ atoms in $|1,\pm 1 \rangle$.
Measuring the atoms in $|1,\pm 1 \rangle$ enables a postselection on experimental realizations with negligible population in level $|h \rangle$, which is an effective postselection on vanishing magnetic-field fluctuations.
While the radio-frequency pulse also reduces the number of atoms in $|2,0 \rangle$, they can still be employed to determine the interferometric phase by a microwave $\pi$/2 pulse on the clock transition $|1,0 \rangle$ to $|2,0 \rangle$.

Figures~\ref{fig:sequence}(c)-(e) show the experimental result for an initial state of $n_0=0.5$ and $\varphi=0$ at $q=1.25 \Omega$ (P$^{\prime}$ phase).
The relative population $n_0$  [Fig.~\ref{fig:sequence}(c)] shows a clear oscillation as it is not affected by magnetic field noise.
In Fig.~\ref{fig:sequence}(d), the relative population after the interferometric sequence, the phase signal, follows the ideal trajectory but picks up substantial noise  after $5-10$~ms.
A postselection on a relative population of $|h \rangle$ versus $|g \rangle$ of less than $35\%$ reduces the fluctuations substantially and collapses the measurements onto the prediction.
However, a mirrored signal appears whenever the magnetic field deviation is large enough for the $|g \rangle$ atoms to cycle once to $|h \rangle$ and back, which is associated with a phase shift of $\pi$.

A meaningful order parameter can be defined to be $|\sin\varphi|$ [Fig.~\ref{fig:sequence}(e)] as measured after a half-period of $n_0$~\cite{supp}.
The experimental data for variable evolution time agree with the expectation.
The mean value of $0.78(3)$ at a half-period allows for a significant discrimination of the P$^{\prime}$ phase versus the adjacent BA$^{\prime}$ phase with an expected order parameter of 0.
The residual noise could be further reduced by a stricter postselection parameter at the expense of prolonged data acquisition.
We note that for both the P$^{\prime}$ and the TF$^{\prime}$ phases, the ideal value of the presented order parameter is 1.
These phases, which are not adjacent anyhow, could nevertheless be distinguished by taking the direction of the evolution into account.
The proposed order parameter together with a time-resolved measurement allows for a discrimination of all three excited-state quantum phases and thus for an experimental verification of the complete excited-state quantum phase diagram in Fig.~\ref{fig:abstract}(a).

We employ the developed order parameter to map out the phase diagram along seven different paths that are shown as colored lines in Fig.~\ref{fig:orderparameterdiagram}(a).
The paths are conceptually different, as they exploit the variation of three different experimental parameters: the QZE $q$, the initial relative population $n_0$, and the initial relative phase $\varphi$.
At fixed QZE, the latter two correspond to a variation of the excitation energy per particle
\begin{align}\label{eq:eta}
    \eta&=\frac{\langle \hat H\rangle}{\Omega N} - \eta_0\\ \notag
    &=\frac{q}{\Omega}\left(1-n_0\right)-2n_0(1-n_0)\textrm{cos}^2\varphi+\frac{1}{2}\left( \frac{q}{2\Omega} -1 \right)^2 \textrm{,}
\end{align}
where the last term corresponds to the ground-state energy per particle $\eta_0$.

First, we vary the QZE $q$ while maintaining an initial state with $n_0=0.5$ and $\varphi=0$ [Fig.~\ref{fig:orderparameterdiagram}(b)].
The order parameter shows the expected behavior; it exhibits large values close to 1 in the TF$^{\prime}$ and the P$^{\prime}$ phase, and small values close to 0 in the BA$^{\prime}$ phase, with sharp ESQPTs in between.
The error bars quantify the stability of the results with respect to a variation of the postselection parameter.
It is varied within a range of 0\%-100\% maximum relative population of $|h \rangle$ and usually fixed to 35\%.
The ESQPTs are broadened by technical fluctuations. Magnetic field detuning changes the value of $q$, fluctuations of the total atom number vary $\Omega$ and preparation imperfections result in deviations of the excitation energies.
The final result is also displayed in the phase diagram Fig.~\ref{fig:orderparameterdiagram}(a), which also includes corresponding measurements for different values of $n_0$.

Second, we vary the excitation energy $\eta$ by preparing initial states with different relative population $n_0$, while keeping the QZE $q$ and initial phase $\varphi$ constant [Fig.~\ref{fig:orderparameterdiagram}(c)].
The results for $\eta<0.05$ are inconclusive, as a low relative population $n_0$ of a few percent makes the phase estimation unreliable. 
This is also indicated by the instability with respect to the postselection parameter.
However, the majority of data for $\eta>0.05$ confirm the ESQPT.

Finally, we vary the excitation energy $\eta$ by adjusting the initial phase $\varphi$ at the relative population $n_0$ of the ground-state stationary point~\cite{supp}, as shown in Fig.~\ref{fig:orderparameterdiagram}(d).
The phase measurement is now evaluated after a quarter-period of the $n_0$-evolution to approximate the presented order parameter with good contrast.
The data show a clear transition from the BA$^{\prime}$ to the P$^{\prime}$ phase for increasing energy.
The results presented in Figs.~\ref{fig:orderparameterdiagram}(c)-(d) appear as vertical lines in the phase diagram Fig.~\ref{fig:orderparameterdiagram}(a).
Together, the evaluation of the order parameter enables a precise determination of the excited-state quantum phases and their transition lines, which presents the main result of this Letter.

In conclusion, we have experimentally mapped out an excited-state phase diagram, a task that up to now remained elusive in any other physical platform, showing that spinor Bose-Einstein condensates constitute a well controllable system for the study of excited-state phases and phase transitions.  
Employing an atomic spin-1 Bose-Einstein condensate, excited quantum states were prepared with a controlled well-defined energy.
The phase diagram was then mapped out by determining an interferometric order parameter, introducing a robust protocol that is designed to be largely insensitive to magnetic field fluctuations.
Our experiments probe a crucial feature of excited-state transitions: they can be crossed not only by quenching the control parameter (in our case the quadratic Zeeman effect), as ground-state transitions, but also by a controllable precise change in the excitation energy, which acts as a second control parameter.  
The probed abrupt change in the qualitative nature of the excited states at a critical excitation energy, experimentally extends the powerful concept of quantum phases to the entire Hilbert space of the spinor quantum gas.\\[0.3cm]

We thank A. Smerzi, L. Pezz\`e, and M. Gessner for inspiring discussions.
We acknowledge financial support from the Deutsche Forschungsgemeinschaft (DFG, German Research Foundation) -- Project-ID 274200144 -- SFB 1227 DQ-mat within the project A02 and under Germany’s Excellence Strategy -- EXC-2123 QuantumFrontiers -- 390837967.
F.A. and B.M.-H. acknowledge support from the Hannover School for Nanotechnology (HSN).
P.F. acknowledges support from the Canada First Research Excellence Fund, Quantum Materials and Future Technologies Program.

B.M.-H. and F.A. performed the experiments.
B.M.-H., F.A. and P.F. analyzed the data.
C.K. and L.S. conceived and designed the experiments with all authors contributing.
The manuscript was written by all authors.

\pagebreak
\clearpage
\onecolumngrid{
\begin{center}
\textbf{\large Excited-State Phase Diagram of a Ferromagnetic Quantum Gas\\(Supplemental Material)}
\end{center}
}
\vspace{1cm}
\setcounter{equation}{0}
\setcounter{figure}{0}
\setcounter{table}{0}
\makeatletter
\renewcommand{\theHequation}{S\arabic{equation}}
\renewcommand{\theequation}{S\arabic{equation}}
\renewcommand{\thefigure}{S\arabic{figure}}
\renewcommand{\theHfigure}{S\arabic{figure}}

\twocolumngrid

\section{Iterative Measurement Method}
To avoid the effect of magnetic-field induced dephasing, we apply an iterative method to track trajectories on the multi-particle Bloch sphere.
Magnetic field fluctuations lead to a coupling between $|g\rangle$ and $|h\rangle$.
The time evolution of the quantum state remains unchanged, as the Hamiltonian does not discriminate between $|g\rangle$ and $|h\rangle$, but the phase read-out by the radio-frequency pulse is strongly affected because it couples to $|g\rangle$ only.
Consequently, we observe a dephasing of the measurement results for more than $4\mbox{--}6$~ms.
The iterative method involves measurements of relative population and phase after an evolution time where the dephasing remains negligible.
After several identical measurement repetitions, the relative population $n_0$ and phase $\varphi$ are sufficiently well determined to allow for a subsequent experimental initialisation of the system at the measured parameters.
Starting with the initial probe state, the procedure is iterated to acquire the full trajectory.
Figure 2 in the main text shows that the iterative method yields trajectories that are well reproduced by our model.
However, the method is based on a reinitialisation with clean coherent states, which neglects the squeezing effects in close vicinity of the separatrix.
Furthermore, the analysis requires a large number of experimental realisations for an accurate estimation of number and phase for all time steps, rendering an exploration of the full phase diagram impractical.
In the following section, we thus report on a more efficient magnetic-field insensitive method.

\section{Magnetic-field Insensitive Measurements}
\begin{figure}[t]
\centering
  \includegraphics[width=\columnwidth]{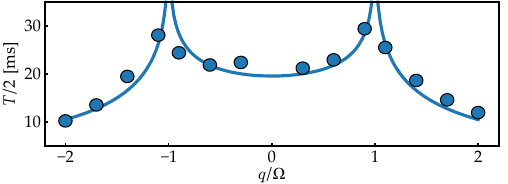}
    \caption{Measurement of the oscillation period.
    The half oscillation period is extracted from parabolic fits at the turning points of the relative population as a function of time.
    The result of the fitting procedure (dots) agrees well with our model (solid line), where the interaction strength $\Omega$ is determined by an independent measurement.
  \label{fig:period}
  }
\end{figure}

\begin{figure}[t]
\centering
  \includegraphics[width=\columnwidth]{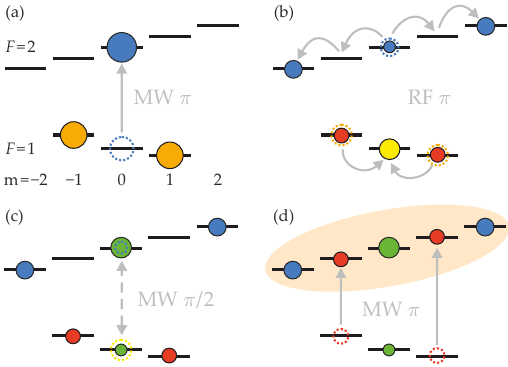}
    \caption{Illustration of the magnetic-field insensitive measurement protocol. 
    With this protocol, the order parameter can be extracted (see text).
    Dashed circles indicate the previous state of the atoms before radiofrequency (RF) or microwave (MW) pulses.
  \label{fig:scheme}
  }
\end{figure}

Our magnetic-field insensitive measurement protocol relies on an individual addressing of the symmetric superposition $|g\rangle$, while using the number of atoms in the antisymmetric superposition $|h\rangle$ for post-correction.
The relative population $n_0$ follows a magnetic-field insensitive oscillation, such that the oscillation period can be measured experimentally, and is well predicted by our model [Fig.~\ref{fig:period}].
The prediction is employed to perform phase measurements at half the oscillation period, where the relative population $n_0$ is recorded.
For the measurement of the phase $\varphi$, the atoms in $|1,0\rangle$ are first transferred to $|2,0\rangle$ [Fig.~\ref{fig:scheme}(a)].
A subsequent radio-frequency $\pi$ pulse transfers the population of the symmetric superposition $|g\rangle$ to the level $|1,0\rangle$, while the antisymmetric superposition $|h\rangle$ remains in the levels $|1,\pm 1\rangle$ [Fig.~\ref{fig:scheme}(b)].
Due to its linear polarisation, the radio-frequency is also resonant with the $F=2$ manifold, transferring $37.5\%$ to $|2,\pm 2\rangle$ each and causing a phase shift of $\pi$ on all $|2,m\rangle$ states~\cite{schafer2014sup}.
As the quadratic Zeeman effect leads to a small deviation from these ideal values, we calibrated these transfer efficiencies $\nu_2=35\%$ and $\nu_{-2}=36\%$ and the residual population $\nu_0=1-\nu_2 - \nu_{-2}$ experimentally.
The desired phase relation is now placed between the two clock states $|1,0\rangle$ and $|2,0\rangle$, which can be rotated into a number difference by a microwave (MW) $\pi/2$ pulse [Fig.~\ref{fig:scheme}(c)].
Before a state-dependent detection, the $|h\rangle$ atoms are transferred to $|2, \pm 1\rangle$ by two MW $\pi$ transitions [Fig.~\ref{fig:scheme}(d)].
From the final number detection of all Zeeman levels in $F=2$, all relevant information is available:
The number of atoms in $|2,\pm 1\rangle$ identifies the original population in $|h\rangle$.
The measured number in $|2,\pm 2\rangle$ allows for an estimation of the original population of $|1,0\rangle$.
From the prior measurement of the relative population, the number of atoms in $|g\rangle$ and equivalently the total number can be obtained.
These measurements fix the occupation of the clock levels before the MW $\pi/2$ pulse, and thus render the measurement of $|2,0\rangle$ after the MW pulse an effective measurement of the phase $\varphi$.
Taking the atoms in $|h\rangle$ into account, this phase measurement can be exploited as an order parameter for the given excited-state phase diagram.
In the next section, we derive how the phase is computed from the measured atom numbers.

\section{Extracting the Order Parameter from Experimental Data}
The order parameter $|\sin \varphi|$ can be extracted from the final atom numbers $N_m$ in the Zeeman sublevels $|2,m\rangle$ of the $F=2$ manifold.
After the time evolution and thus at the start of our measurement protocol (see previous section), the state of the sBEC is well approximated by a product of single-particle states
\begin{equation}
	|\psi\rangle = \sqrt{n_0}\textrm{e}^{-i\varphi}|1,0\rangle+\sqrt{1-n_0}(\cos\beta\,|g\rangle -i\sin\beta \,|h\rangle),
\end{equation}
where $\beta$ represents an unknown mixing angle between $|g\rangle$ and $|h\rangle$ due to magnetic-field fluctuations and $\varphi$ is the phase that we would like to extract.
Our measurement protocol yields
\begin{equation}\label{eq:npm2}
	N_{\pm 2} = n_0 \nu_{\pm2} N~~~~~~~~~~~ \Rightarrow ~~~~~~ N=\frac{N_2+N_{-2}}{n_0 (\nu_2 + \nu_{-2})},
\end{equation}
\begin{equation}\label{eq:npm1}
	N_{\pm 1} = \frac{1-n_0}{2}\sin^2\!\beta\, N~~ \Rightarrow ~~ \sin^2\!\beta = \frac{N_1+N_{-1}}{(1-n_0)N},
\end{equation}
and
\begin{align}\label{eq:n0}
	N_0 &= \frac{1}{2}\left(\nu_0n_0+(1-n_0)\cos^2\!\beta\right)N \\ \nonumber
	&+\sqrt{\nu_0n_0(1-n_0)}\cos\beta \sin(\varphi-\alpha_0) N.
\end{align}
Here, $\nu_m$ and $\alpha_m$ describe the effect on the atoms in $|2,0\rangle$ when separating $|g\rangle$ from $|h\rangle$ via a radio-frequency pulse (see previous section).
Inserting the expression for $\sin^2\!\beta$ from Eq.~\ref{eq:npm1} into Eq.~\ref{eq:n0}, we obtain
\begin{equation}\label{eq:OP}
	|\sin\varphi|=\frac{\left|\left(N_0+\frac{1}{2}(N_1+N_{-1})\right)/N+\frac{1}{2}\left((1-\nu_0)n_0-1\right)\right|}{\sqrt{\nu_0n_0(1-n_0-(N_1+N_{-1})/N)}}.
\end{equation}
Note that we can extract only the absolute value of $\sin\varphi$ because we do not have access to the sign of $\cos\beta$.
Thus, Eq.~\ref{eq:OP} together with the expression for $N$ from Eq.~\ref{eq:npm2} can be directly used to evaluate the order parameter $|\sin \varphi|$ as shown in Fig.~4(b-d) in the main text.\\

\begin{figure*}[t]
\centering
  \includegraphics[width=\textwidth]{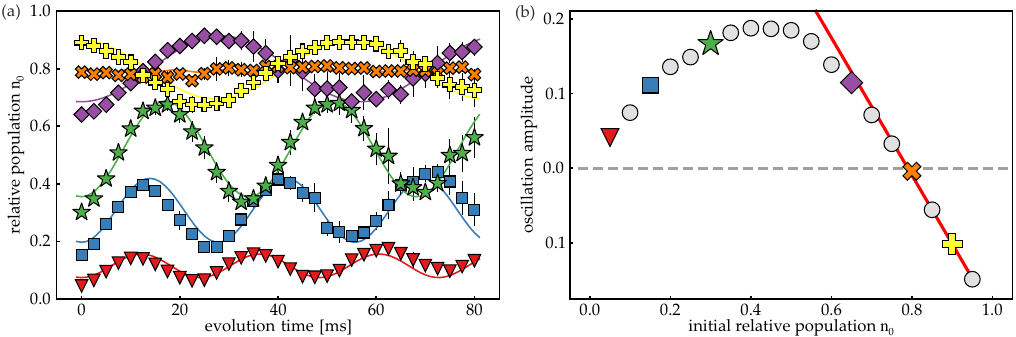}
    \caption{Determination of a stationary point.
    Initial parameters are $q/\Omega=1.25$ and $\varphi=0$.
    (a) For different initial values, the measured relative populations $n_0$ follow an oscillatory behavior (symbols).
    The data are fitted with sinusoidal functions (solid lines) to extract the amplitude.
    Error bars present the standard deviation.
    (b) The fitted oscillation amplitude as a function of the initial population shows a characteristic zero crossing which identifies the stationary point at $n_0=0.79$.
    Coloured symbols refer to the data in panel (a).
    The solid line is a linear fit near the zero crossing.
  \label{fig:populationoscillation}
  }
\end{figure*}

\section{Determination of Stationary Points}
The phase transition by a variation of the initial phase $\varphi$ is best explored when initializing the system at the relative population of the stationary point [Fig.~4(d) in the main text].
In this way, the variation occurs as orthogonal to the trajectories as possible and thus the contrast after a quarter-period is favourable.
As an example, we determine a stationary point at $q/\Omega=1.25$, $\varphi=0$ and varying $n_0$.
The resulting population oscillations are presented in Fig.~\ref{fig:populationoscillation}(a).
The value of $n_0=0.79$, where zero oscillation amplitude is expected (and the sign changes) corresponds to the stationary point [Fig.~\ref{fig:populationoscillation}(b)].
This stationary point determination serves as the starting point of the characterisation of the phase transition in Fig.~4(d) in the main text.

\end{document}